\documentclass{elsart1p}

\usepackage{amssymb}
\usepackage{graphicx}
\usepackage[utf8]{inputenc}
\begin{document}

\begin{frontmatter}

\title{Spatial Patterns of Wind Speed Distributions in Switzerland}


\author{Mohamed Laib         \and
        Mikhail Kanevski 
}



\author{}

\address{Institute of Earth Surface Dynamics, \\
          Faculty of Geosciences and Environment,\\ 
          University of Lausanne, CH1015 Lausanne, Switzerland\\ 
          Email: mohamed.laib@unil.ch}

\begin{abstract}
This paper presents an initial exploration of high frequency records of extreme wind speed in two steps. The first consists in finding the suitable extreme distribution for $120$ measuring stations in Switzerland, by comparing three known distributions: Weibull, Gamma, and Generalized extreme value. This comparison serves as a basis for the second step which applies a spatial modelling by using Extreme Learning Machine. The aim is to model distribution parameters by employing a high dimensional input space of topographical information. The knowledge of probability distribution gives a comprehensive information and a global overview of wind phenomena.
Through this study, a flexible and a simple modelling approach is presented, which can be generalized to almost extreme environmental data for risk assessment and to model renewable energy.

\end{abstract}


\end{frontmatter}



\section{Introduction}
\label{intro}

Wind can be regarded as either a positive or negative phenomenon. The positive aspect is the renewable energy it produces which has encouraged the Swiss Federation to expand the proportion of power produced by wind speed \cite{fcspwind}. On the other hand, enormous losses have been caused by extremely violent wind-storms in the country \cite{wind_economic_impact,windthrow_swiss}, an excellent catalogue of which has been produced by Stucki et al. 2014 \cite{wind1859_intro1}, and Usbeck et al.\cite{storm_swiss}.\\
The first wind energy facility in Switzerland was started in 1986 with an energy output of 28 kilowatts. According to the Swiss Federation reports in 2015, there are 34 wind power plants which produce around 110 gigawatts of electricity. The largest wind park is on Mont Crosin in the Bernese Jura. This facility comprises 16 wind turbines with a total output of 29.2 megawatts \cite{httpbfe}. To improve the use of this environmental source, a well-developed statistical field for this type of analysis has been proposed. Most methods used to analyse wind data deal with semi-parametric approaches begin by finding the best probability distribution and then confirm results with parametric and non-parametric tools.\\
Since wind data is known for the presence of extremes, the modelling of these data requires extreme probability distribution in order to study the behaviour of tail in data.
Extreme value theory (EVT) has been the most frequently applied modelling approach. The main purpose is to find estimators of the suitable distribution for the studied data \cite{weibull,galambos,castillo}.\\
There are many areas where EVT plays an indispensable role for modelling rare events, such as environmental risk (wind, temperature, rainfall, etc.)\cite{wind_evt1,wind_evt2_china,wind_evt3,wind_evt4_copulas}. Numerous existing parametric and non-parametric estimation methods are used to find estimators. This work uses the maximum likelihood as estimation method.
Besides the method mentioned above, machine learning algorithms (MLA)\cite{vapnik98,tibshiranB} are rapidly gaining popularity in modelling environmental phenomena\cite{KPT2009,Leuenberger_Dec15,wind_ML_fuzz}. Machine learning is a part of artificial intelligence. Whose objective is to find non-linear dependencies observed in data, and to understand better the structure between the input and the output.

Several papers propose different approaches to make use of the performance of machine learning in wind speed modelling. One use more information such as environmental data \cite{winden}, therefore the parameters of the proposed extreme probability distribution are modelled by using other environmental variables as input data. Many algorithms are used for this purpose notably random forest \cite{uk_wind}. A comparison between extreme learning machine, support vector machine, and artificial neural networks has been carried out \cite{comparison,appelm}. This comparison favours Extreme Learning Machine for its quality of modelling, its rapidity and simplicity.
This study uses the Extreme Learning Machine (ELM) proposed by Huang et al., $2006$. Its structure is similar to that of a classical multilayer perceptron (MLP). Moreover, ELM has one parameter to optimise which is the number of hidden nodes. This parameter makes ELM easy to apply and to control the complexity of the phenomenon under study.
Its main advantage  is the speed of the training step, and the capacity to learn complex data. The techniques of cross-validation and data splitting help to avoid overfitting, and also to test the accuracy of the model \cite{tibshiranB}. Furthermore, the use of ELM requires a consistent methodology to take into account the randomness in generating the weights.
 
The aim of this study is to find a flexible approach to understand the behaviour of extreme wind speed in Switzerland. The proposed approach combines two steps: The first uses a parametric method called maximum likelihood, to estimate probability distribution parameters. The selection of the best extreme distribution is based on the Kolmogorov-Smirnov test and the quantile-quantile plot. These two tools are commonly used to compare if the data are well modelled by the proposed probability distribution.

The second step of this work applies the Extreme Learning Machine to model the estimated parameters of the first step.

The main results are presented as probability maps, and parameters are also mapped to visualise them. In addition, the ELM results are quantified to show their efficiency to model different distribution parameters.
All of these steps are carried out by using $extRemes$ and $elmNN$ packages of the R language \cite{lanR}.

This paper is organized as follows: Section $2$ presents an exploratory analysis of the used data. Section $3$ explains the first step which consists in finding the suitable distribution. Section $4$ gives a brief introduction to ELM, and the proposed spatial modelling. In section $5$, the main results are discussed, and in the last section the conclusions are given with suggestions for future research.

\section{Study Area and Dataset}
\label{sec:2}
\subsection{Study Area}
\label{sec:2.1}
\begin{figure}
\centering
\includegraphics[scale=0.7]{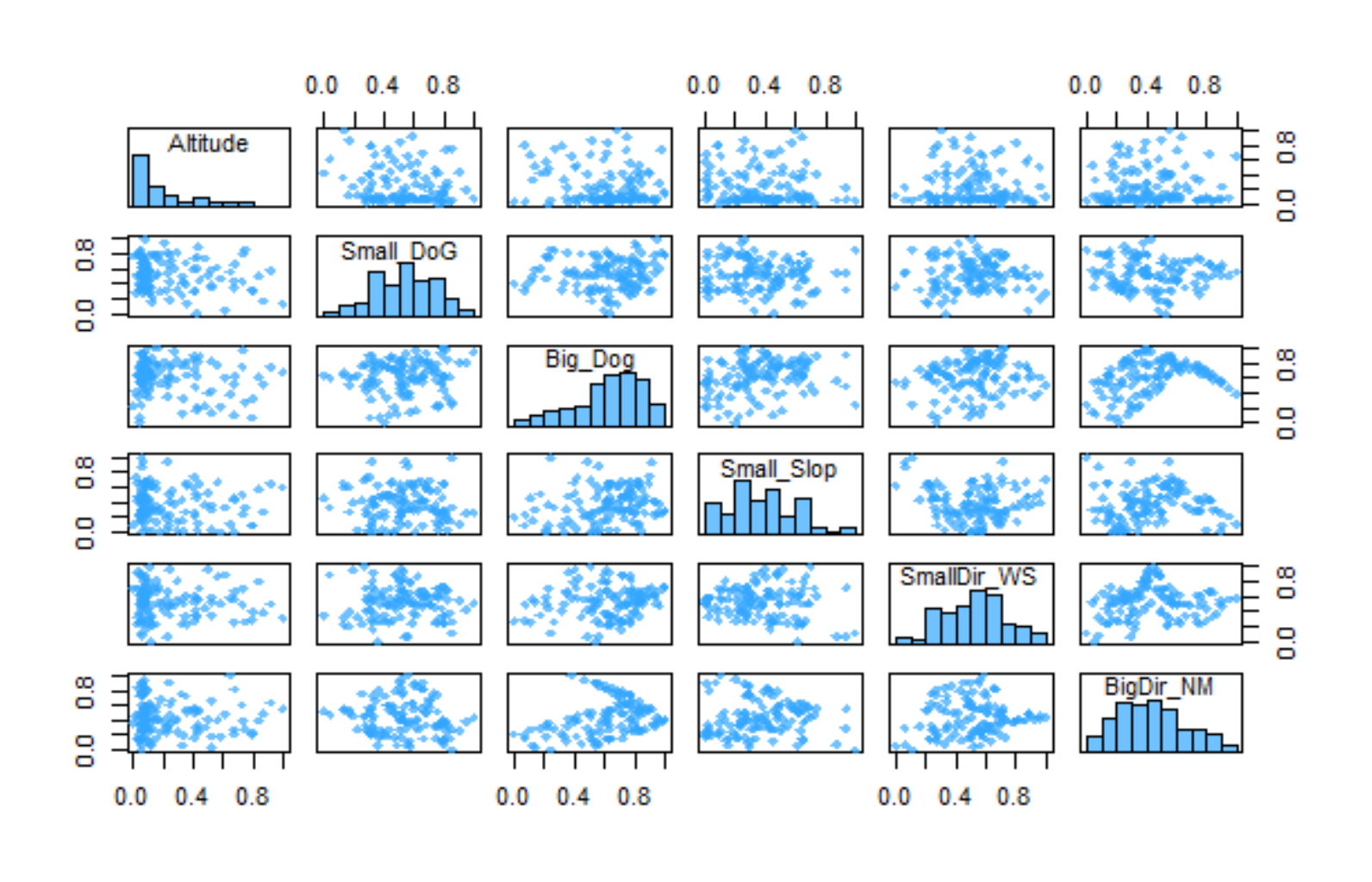}
\caption{Scatter-plot of some variables from the input space used for training ELM scaled in the [0,1] interval.}
\label{fig:1}  
\end{figure}

This study was performed in Switzerland, which has a total area of $41,285$ $km^2$ with three basic topographical area: the Jura mountain on the west, the central plateau, and the Swiss Alps to the south which comprise almost all the highest mountains of the Alps. The altitudes varie between:  $198$ $m$ in canton Ticino, to  $4634$ $m$ in canton Valais \cite{wikiswiss}. All this information is summarized in an input space of thirteen variables, including coordinates $(X,Y)$ at $250$ $m$ resolution (Table 1, see details and descriptions in \cite{Robert}). These variables are used to model parameters provided from each measuring stations. Fig.\ref{fig:1} shows some of variables used as input space. 
\begin{center}
\begin{table}[h]
\caption{Input space variables generated from digital elevation model.}
\begin{tabular}{|l|l|l|}
  \hline
ID & Name of the variable  & Scale \\
\hline
$X$ 	     & $\:$  X coordinate & $\qquad$ \\
  $Y$	       &  $\:$  Y coordinate & $\qquad$ \\
  $Z$ 	       & $\:$  Z (elevation) & $\qquad$ \\
  $dog_{s}$	    & $\:$  Diff. of Gauss. at small scale & $\:   \sigma_{1}=0.25 \;$ km $\quad \sigma_{2}=0.5 \;$ km   \\
  $dog_{m}$	    & $\:$  Diff. of Gauss. at medium scale &  $\:  \sigma_{1}=1.75 \;$ km $\quad \sigma_{2}=2.25 \;$ km  \\
  $dog_{l}$	    & $\:$  Diff. of Gauss. at large scale &  $\: \sigma_{1}=3.75 \;$ km $\quad \sigma_{2}=5 \;$ km  \\
  $S{s}$  & $\:$  Slopes at small scale &  $\: \sigma=0.2 \;$ km  \\
  $S{m}$    & $\:$ Slopes at medium scale & $\:  \sigma=1.75 \;$ km  \\
  $S{l}$ 	    & $\:$  Slopes at large scale &  $\: \sigma=3.75 \;$ km  \\
  $d{ns}$   & $\:$  Dir. deriv. in South–North dir. at small scale &  $\: \sigma=0.25 \;$ km  \\
  $d{ws}$    & $\:$  Dir. deriv. in East–West dir. at small scale &  $\: \sigma=0.25 \;$ km  \\
  $d_{nm}$   & $\:$  Dir. deriv. in South–North dir. at medium scale &  $\: \sigma=1.75 \;$ km  \\
  $d_{wm}$    & $\:$ Dir. deriv. in East–West dir. at medium scale &  $\: \sigma=1.75 \;$ km  \\
  \hline
\end{tabular}
\end{table}
\end{center}

\subsection{Wind Data}
\label{sec:2.2}
\begin{figure}
\centering
\includegraphics[scale=0.45]{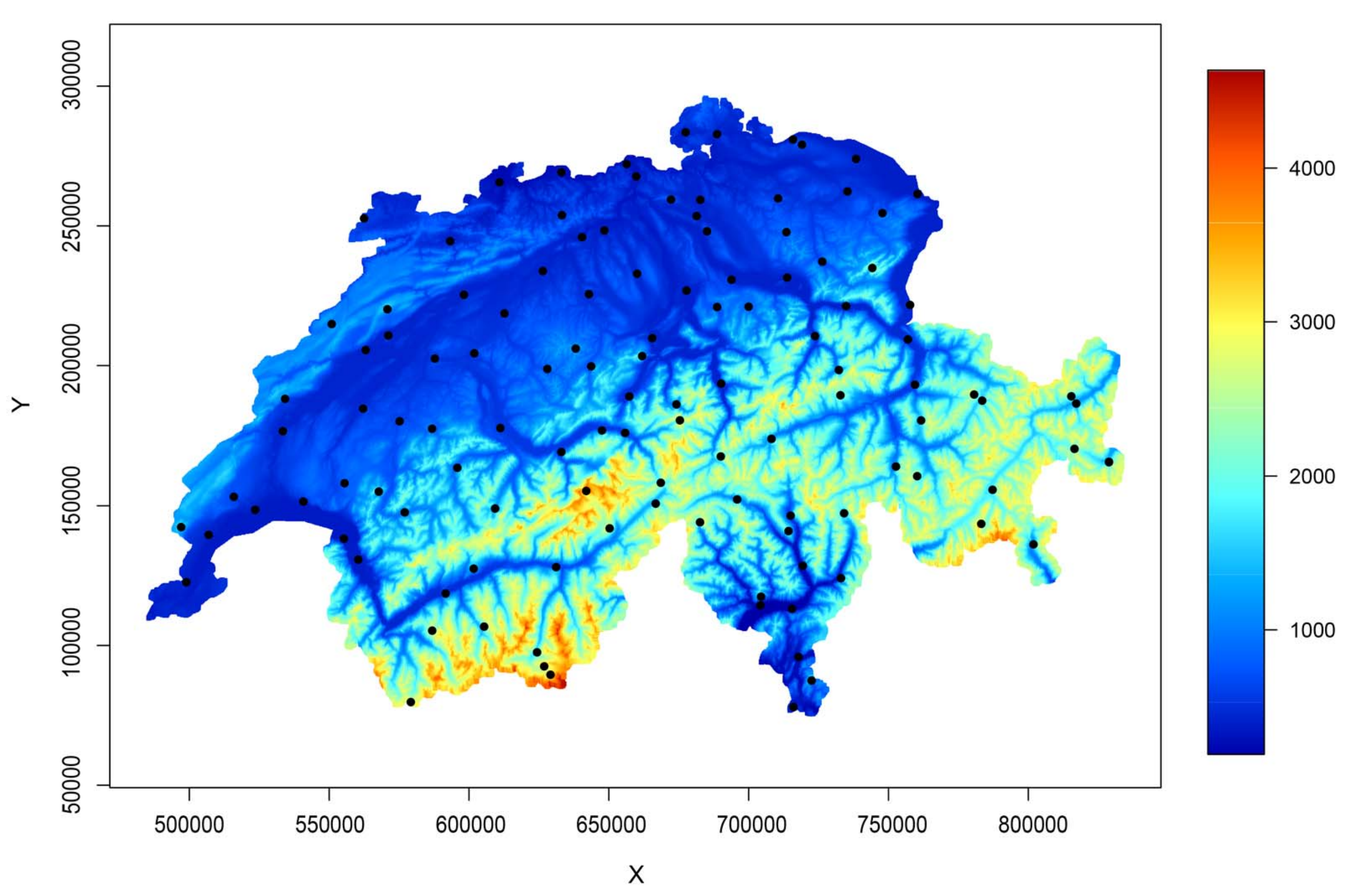}
\caption{Locations of MeteoSwiss stations.}
\label{fig:2}  
\end{figure}

Wind data used in this work were collected from the website of the Federal Office of Meteorology and Climatology of Switzerland (IDAWEB, MeteoSwiss). They present wind speed measurements at weather stations  distributed in all Switzerland (fig.\ref{fig:2}), at different elevations, from $203$ $m$ to $3580$ $m$. In total there are more than $148$ stations. However, some stations were eliminated because they contain an important number of missing values. The final dataset contains measurements of $120$ stations for two years ($2012$ and $2013$), taken at $10$ minutes intervals. This important high frequency allows us to obtain good approximation of wind speed distribution, even the behaviour of extremes.

\begin{figure}
\centering
\includegraphics[scale=0.45]{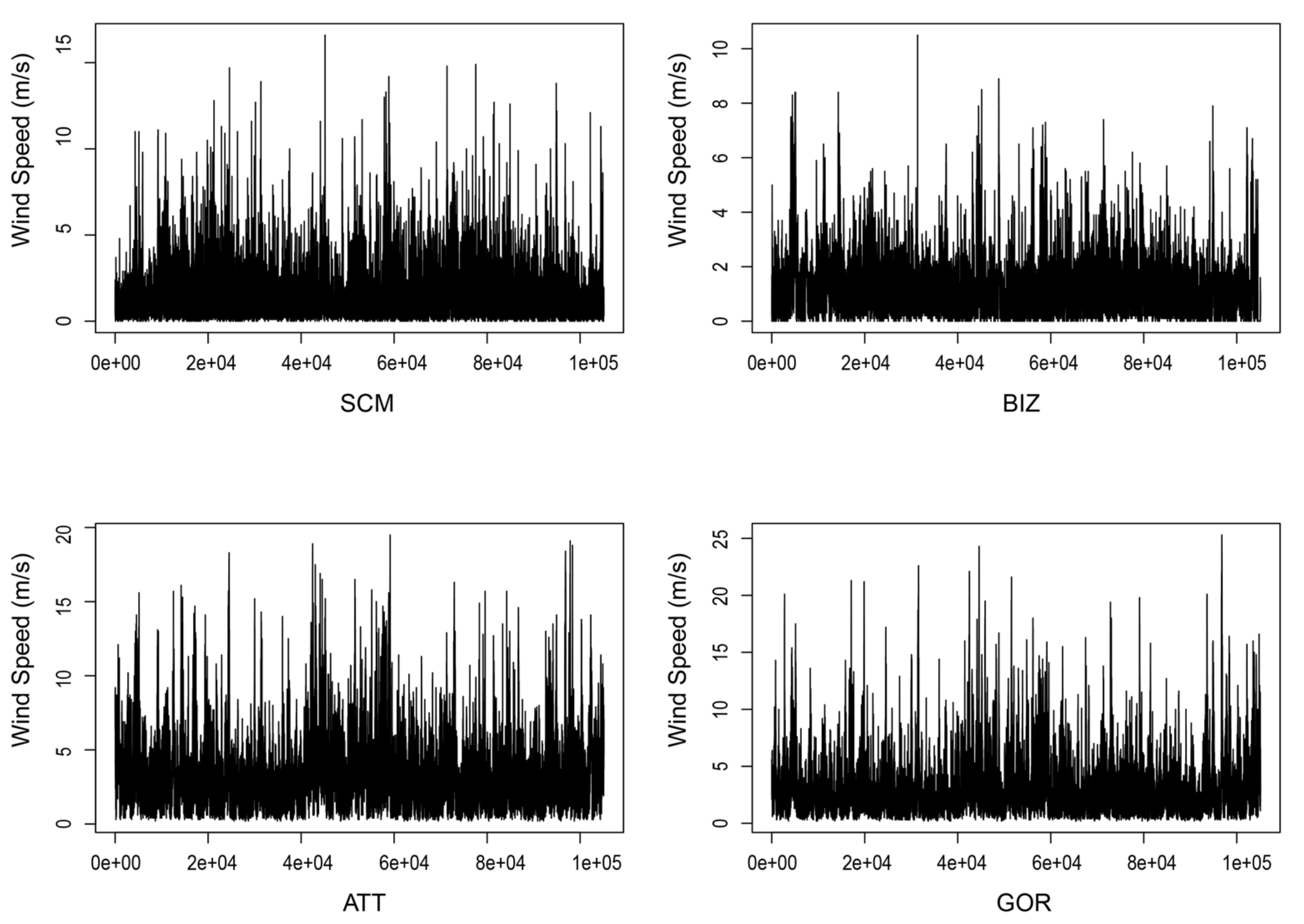}
\caption{Observations of some measuring stations.}
\label{fig:3}  
\end{figure}

\begin{figure}
\centering
\includegraphics[scale=0.45, angle=90]{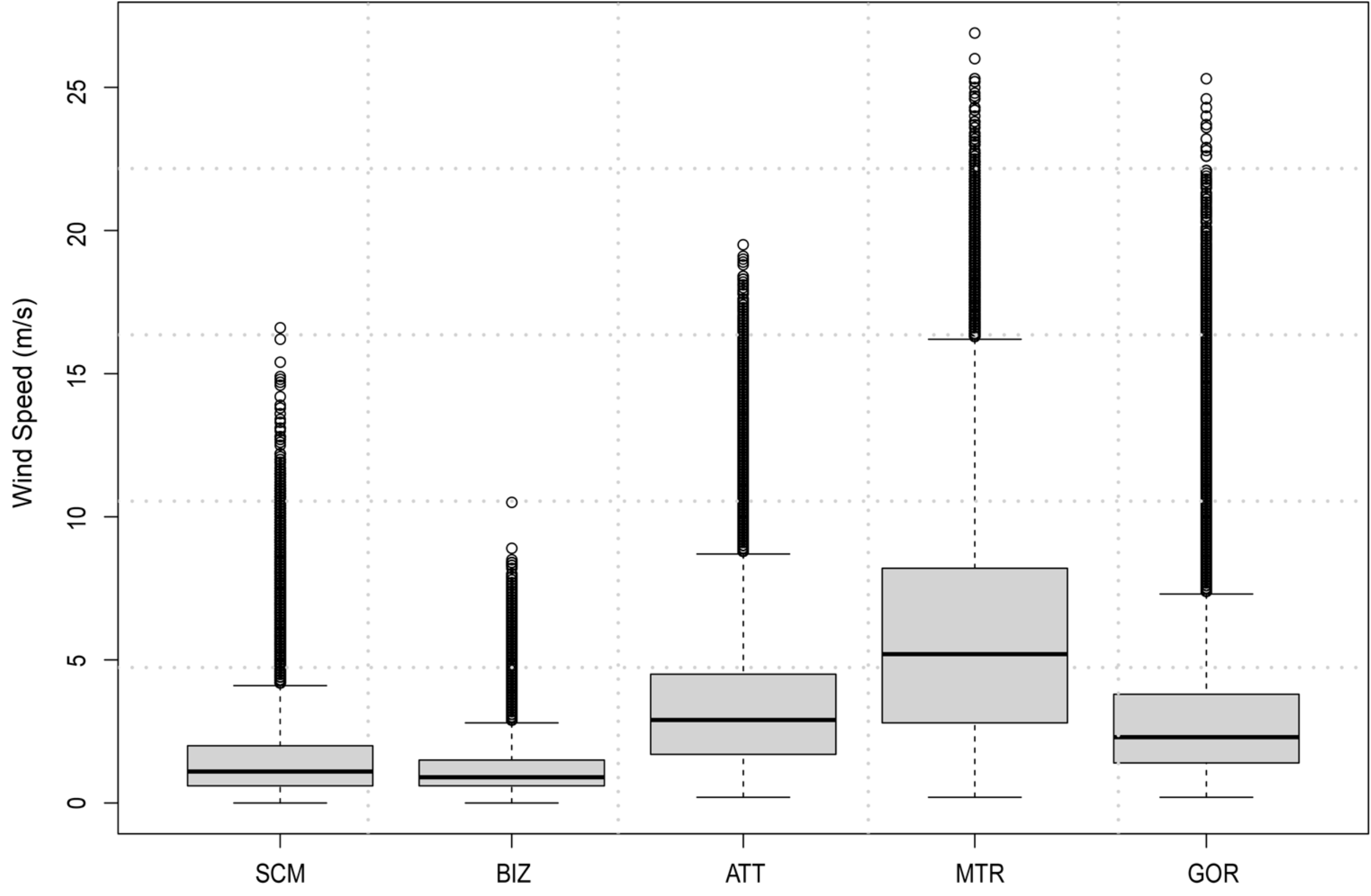}
\caption{Boxplot of some measuring stations.}
\label{fig:4}  
\end{figure}

Fig. \ref{fig:3},\ref{fig:4} show examples of some measuring stations. 
The time series plots do not indicate any significant increasing or decreasing trends. Fig. \ref{fig:4} shows the presence of extreme wind speed, which leads to propose extreme distributions to model the data.

\section{Wind Speed Distribution}
\label{sec:3}
As mentioned above, this work leads off with a parametric estimation using maximum likelihood. \\
Maximum likelihood is a very simple tool to find estimators. It chooses the value of the parameter which maximizes the following likelihood function  \cite{colesB}:\
\begin{equation}
L(\theta)=\prod^{n}_{i=1}f(x_{i};\theta)
\end{equation}
where $x_{i}$ are independent realizations of a random variable with a probability density function $f(x_{i};\theta)$.\
As is known, it is more convenient to work with the log-likelihood function:
\begin{equation}
log \: L(\theta)=\sum^{n}_{i=1} \, log f(x_{i};\theta)
\end{equation}
The log-likelihood takes its maximum at the same point as the likelihood function, and it is found by differentiating the log-likelihood and equating to zero.\\
The parameters of each proposed distribution are found in order to compare.
Several papers link wind data with the following extreme distributions:

\subsection{Weibull Distribution} 
\label{sec:3.1}
Proposed as a wind speed distribution \cite{Epro1}, Weibull distribution is a two parameters distribution, with the following density function \cite{reisx}:

\begin{equation}
f(x;\lambda,k)=\left\{\begin{array}{rcl}
\frac{k}{\lambda} (\frac{x}{\lambda})^{k-1} e^{-(\frac{x}{\lambda})^{k}} \qquad x\geq 0\\
\\
0   \qquad \qquad  \qquad x < 0
\end{array}\right.  
\end{equation}

where $k$ is the shape parameter and $\lambda$ is the scale. This function is a continuous probability distribution and mostly used to describe wind data.

\subsection{Gamma distribution} 
\label{sec:3.2}
Gamma distribution is also a two parameter distribution. It used to present several phenomena especially those that varies over time, and it is defined by the following density function:
\begin{equation}
f(x;\alpha,\beta)=\frac{\beta^{\alpha}x^{\alpha-1}e^{-x\beta}}{\Gamma(\alpha)} \qquad for \; x \geq0 \; and \; \alpha,\beta > 0
\end{equation}

where $\alpha$ is the shape and $\beta$ is the rate \cite{statex}.

\subsection{Generalized Extreme Values}
\label{sec:3.3}
\begin{figure}
\centering
\includegraphics[scale=0.35]{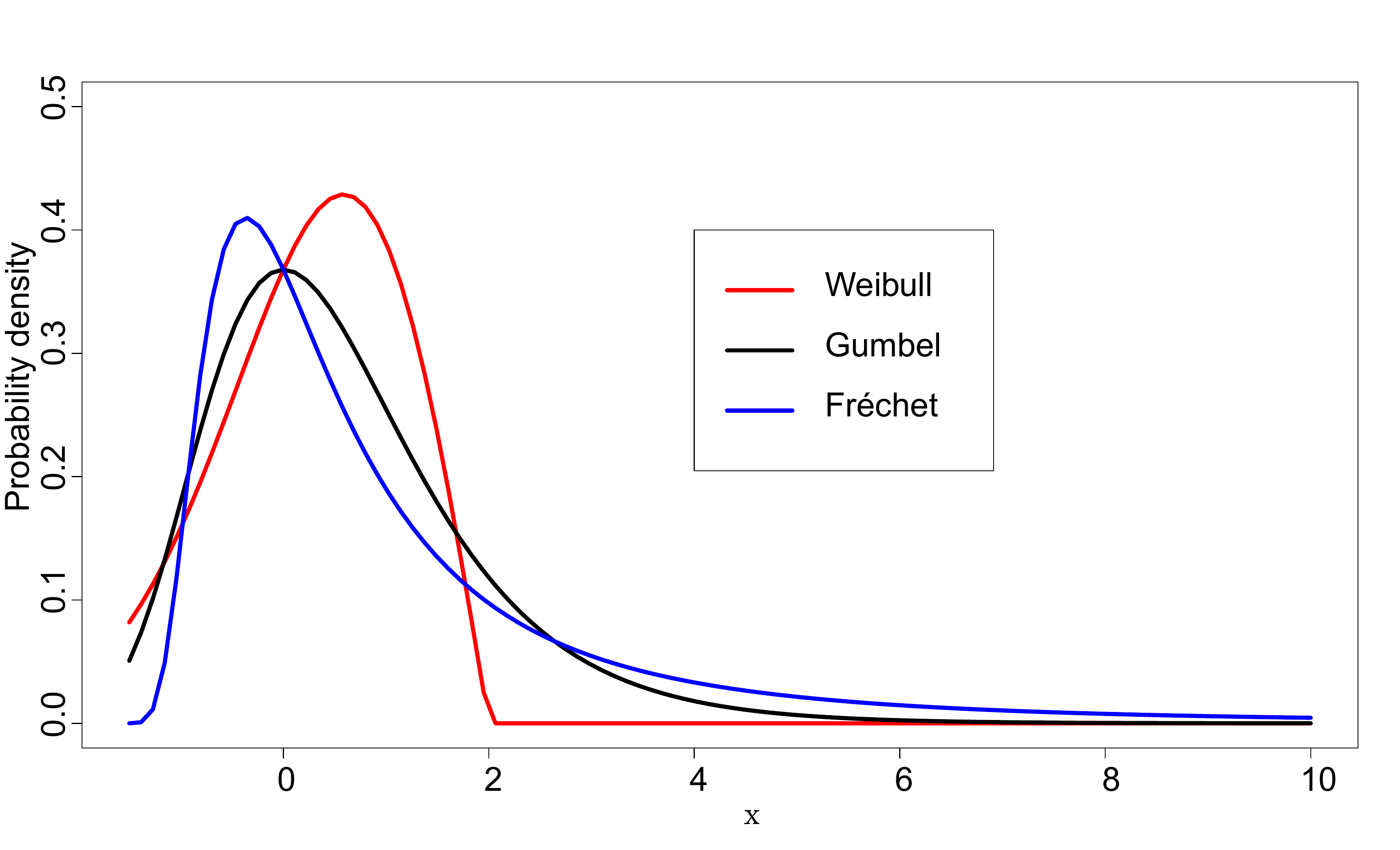}
\caption{The three subfamilies of the GEV with $\mu=0$, $\sigma=1$. And $\xi=-0.5$, $0$, $0.5$ for Weibull, Gumbel, Fréchet respectively.}
\label{fig:5}  
\end{figure}

It combines three families of distributions: Gumbel, Fréchet and Weibull. The GEV distribution is used to model the maxima of long sequences of random variables, and the treatment of risk \cite{colesB}. It is defined as follow:
\begin{equation}
F(x;\mu, \sigma, \xi ) = exp\{ -[1+\xi(\frac{x-\mu}{\sigma})]^{\frac{-1}{\xi}} \} 
\end{equation}

for $ 1+ \xi(x-\mu)/\sigma > 0$, where $\mu $ is the location parameter, $\sigma $ is the scale parameter and $\xi $ the shape. The latter indicates the tail behaviour of the distribution. Fig \ref{fig:5} shows the different forms of the probability density for each subfamilies according to the value of the shape parameter $\xi$, and the subfamilies are defined as following: 
\begin{itemize}
\item Gumbel distribution or type I when $\xi =0$. 
\item Fréchet or type II when $\xi>0 $. 
\item Weibull distribution or type III when $\xi <0 $.
\end{itemize}

The comparison between these distributions is based on the Kolmogorov-Smirnov test, and a graphical method called quantile-quantile plot with a visual statistical protocol, as it is proposed in \cite{protost}.

\section*{Comparison tools}
\label{sec:3.4}

\subsection*{The Kolmogorov-Smirnov goodness of fit test}
\label{sec:3.5}
There are a number of tests to check the goodness of fit for a probability distribution. Among the most used is the Kolmogorov-Smirnov test, which can be applied on continuous distribution. This test is based on the maximum difference of the empirical and the proposed theoretical distribution \cite{KSM}.
The Kolmogorov-Smirnov goodness of test statistic is defined as follows:

\begin{equation}
D= Max \mid F(X_{i})- \frac{i}{N}\mid \qquad for \qquad {1 \leq i \leq N}
\end{equation}

The smaller the value of Kolmogorov-Smirnov statistic is, the better the goodness-of-fits is. In order to confirm the results given by the early goodness of fit, the quantile-quantile plot is proposed.

\subsection*{Quantile-quantile plot}
\label{sec:3.6}
Shortly QQ-plot, which is a graphical method to compare distributions. based on the observation of the quantiles of each distribution. The linearity in the graph is easily verified, or furthermore, it can be quantified by the correlation coefficient \cite{qpEVT}.

\begin{figure}
\centering
\includegraphics[scale=0.3]{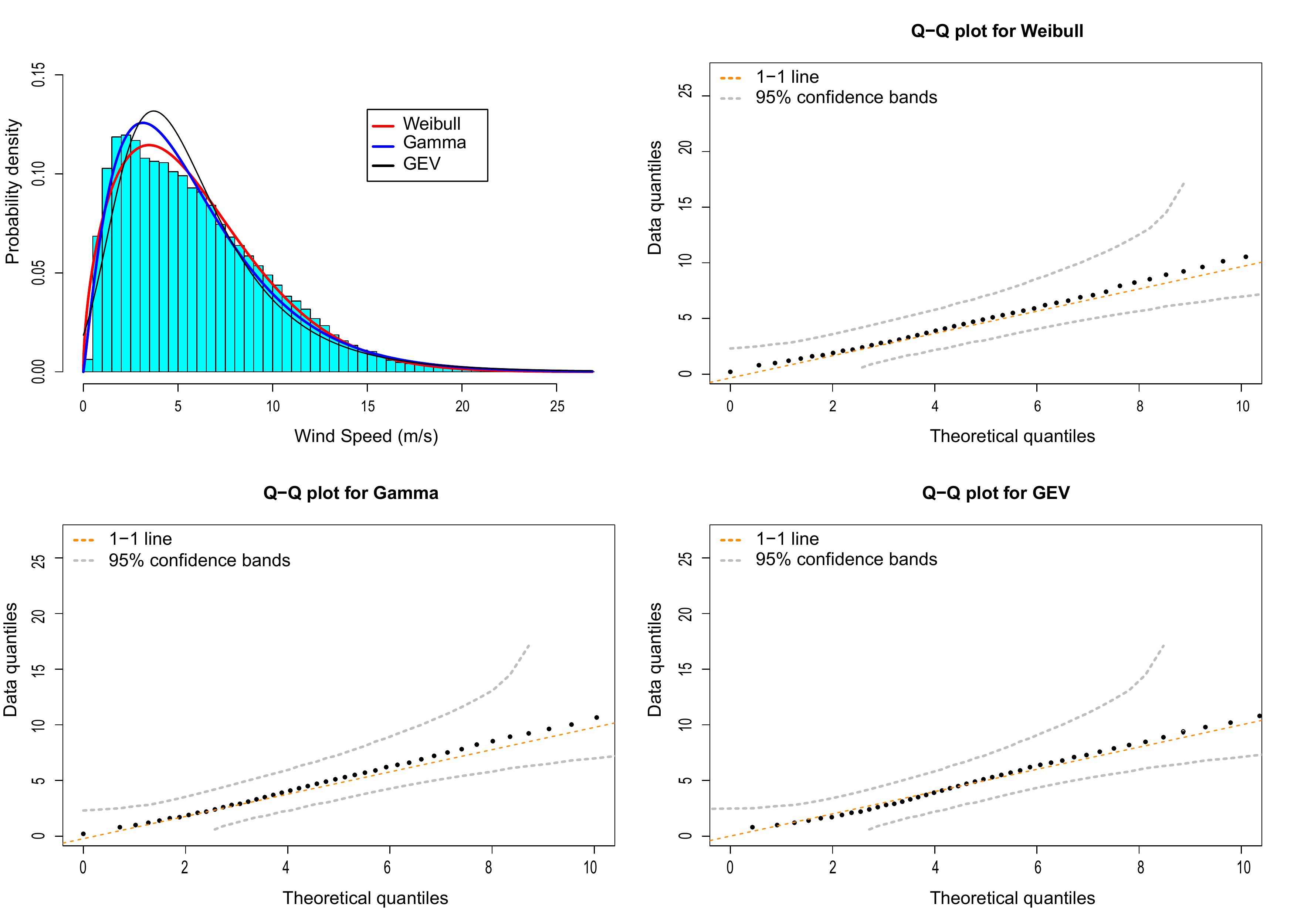}
\caption{Wind Speed at $Matro$ station (2171 meters). The Kolmogorov-Smirnov test gives the following values: $0.1229$, $0.1655$, $0.1102$  for Weibull, Gamma, and Generalized Extreme values respectively.}
\label{fig:6}  
\end{figure}

\begin{figure}
\centering
\includegraphics[scale=0.3]{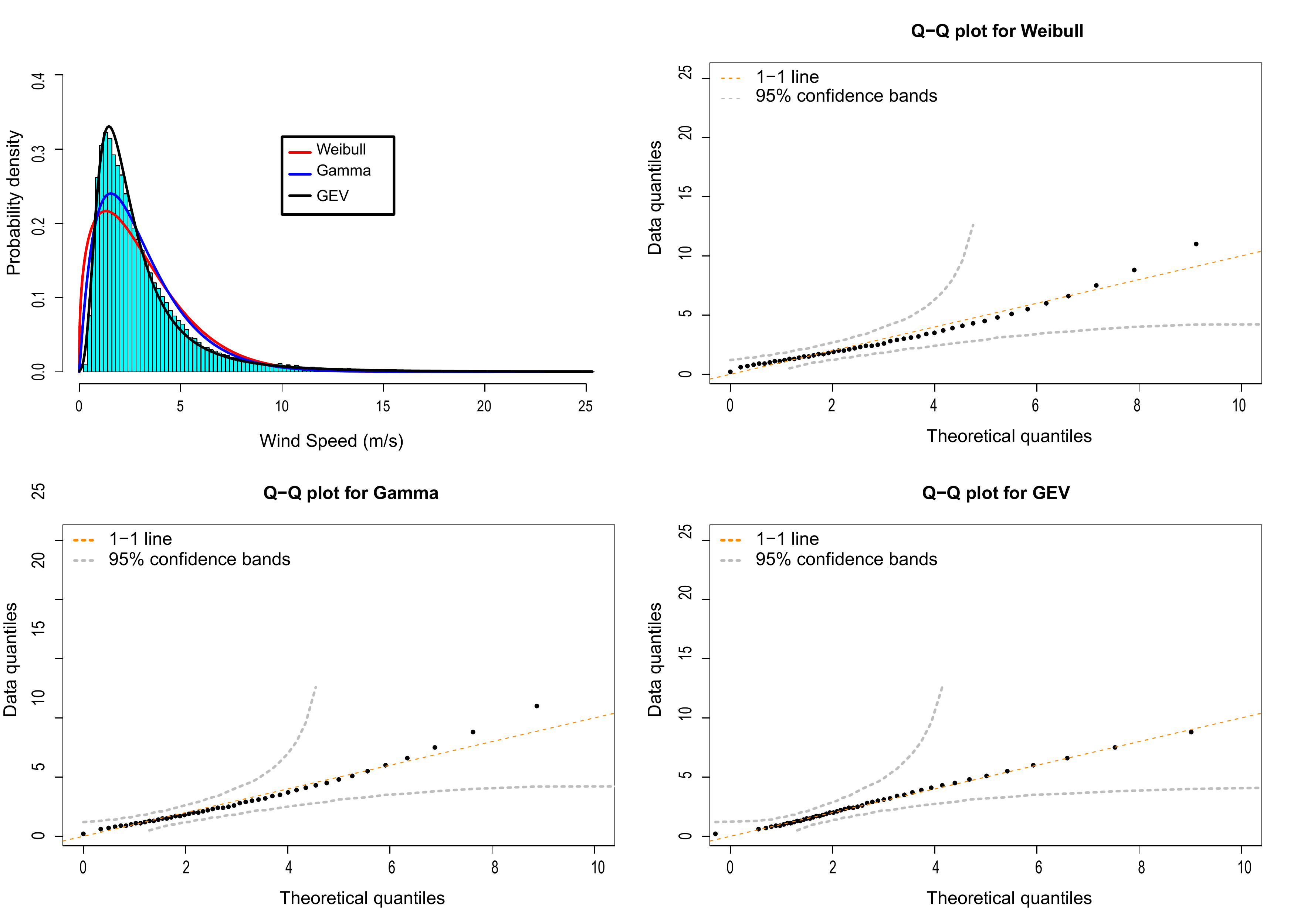}
\caption{Wind Speed at $Gornergrat$ station (3129 meters). The Kolmogorov-Smirnov test gives the following values: $0.1984$, $0.2023$, $0.1026$  for Weibull, Gamma, and Generalized Extreme values respectively.}
\label{fig:7}  
\end{figure}

In this case of study, the data are well-modelled by the Generalized Extreme Value. As described by the QQplot (Figs. 6-7 as examples), extremes are well fitted by the GEV. Furthermore, the Kolmogorov-Smirnov test confirms that the best probability distribution, for the used data, is the GEV  (Fig. 8). According to this comparison, the remain work is based on the GEV. Therefore these parameters ($\mu$, $\sigma$, $\xi$) are modelled by using ELM.

\begin{figure}
\centering
\includegraphics[scale=0.4, angle=90]{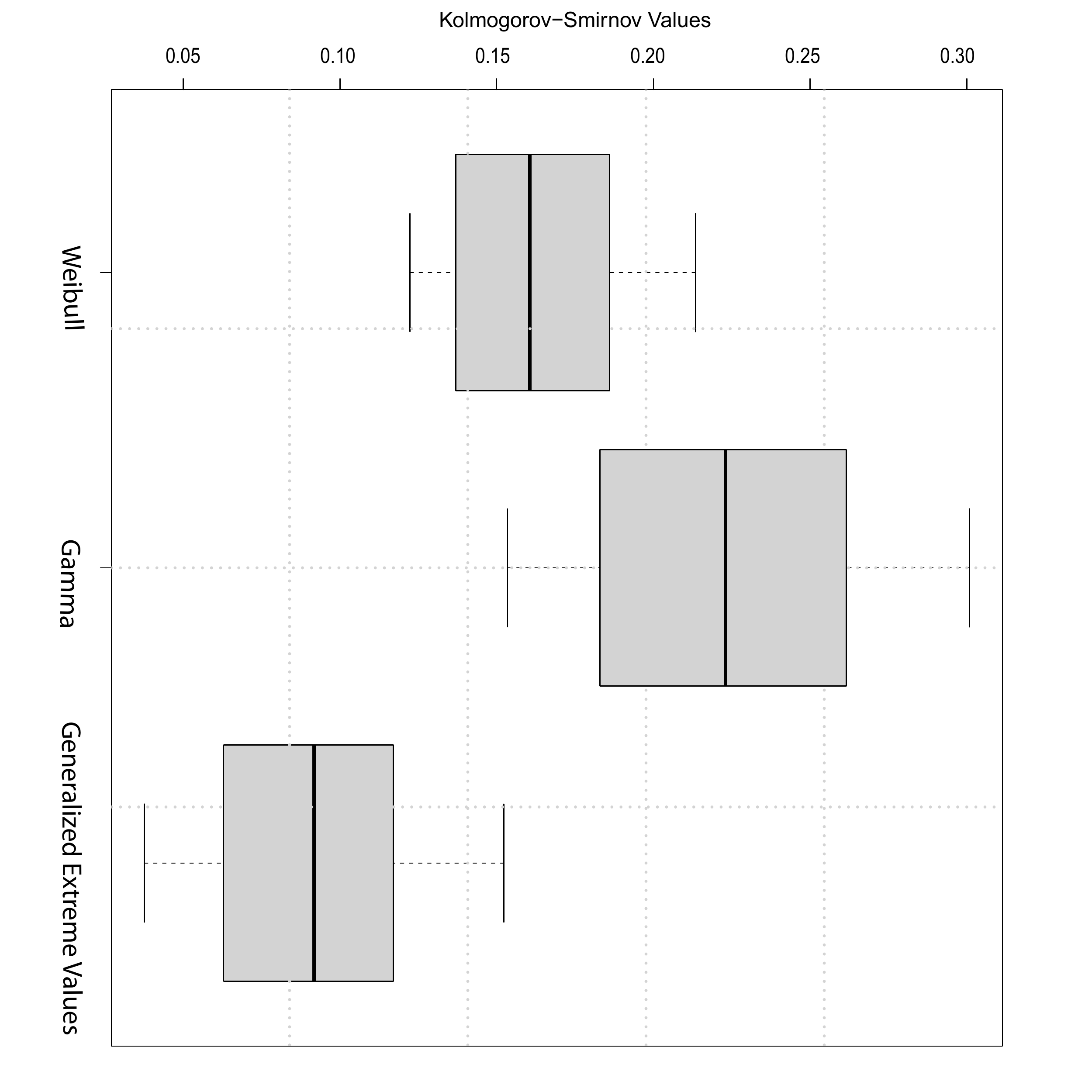}
\caption{Values of the Kolmogorov-Smirnov test for each probability distribution.}
\label{fig:8}  
\end{figure}

\section{Spatial Modelling}
\label{sec:4}
The second step of this study deals with a spatial modelling using Extreme Learning Machine \cite{ELM}.
ELM is inspired by the multi-layer perceptron (MLP) with one hidden layer. For a fixed number of hidden nodes $N$, ELM generates randomly the weights and the biases of each node. Then the result passes through a differentiable activation function $g$ which gives the matrix $H$ where each row corresponds to the output of hidden layer for one input data vector:
\begin{equation}
H_{ij}=g(x_{i}.w_{j}+b_{j})
\end{equation} 
where $x_{i}=x^{1}_{i},x^{2}_{i}, \ldots ,x^{d}_{i} \: (i=1,\ldots, n)$ are the input data, $w_{j} \: (j=1,\ldots,N)$ are the vectors of weights, and $b_{j}$ are the biases of each node.\\
To get the connection vector $\beta$ between the hidden layer and the output layer, ELM uses the Moore-Penrose generalized inverse of the matrix $H$:

$\beta=H^{\dagger} y$ \

These operations give at the end new predicted data points as well as the validation and the testing errors.
And for more efficiency and clarity, the following methodology is used to validate the given model:

\begin{itemize}
\item Data are projected into the interval [0,1], and then are split into training and testing set, in total $30$ measuring stations are assigned as testing set and $90$ as training set.
\item The remain data are used to train ELM with N number of hidden nodes where $N \in \{1, \ldots, 100\} $
\item The optimal number of nodes $N$ is selected by using k-fold cross-validation with respect to the mean square error ($k=6$).
\item Then the optimal model is generated and evaluated according to the mean square error for the testing set.
\item This process of learning is repeated $20$ times, with random splitting of the data, and at the end the mean of repetition is taken.
\end{itemize}

The same process is carried out for all parameters of the GEV.\\
Fig.9 shows that the estimated parameters by maximum likelihood are adequate with the predicted by ELM models.

\begin{figure}
\centering
\includegraphics[scale=0.24]{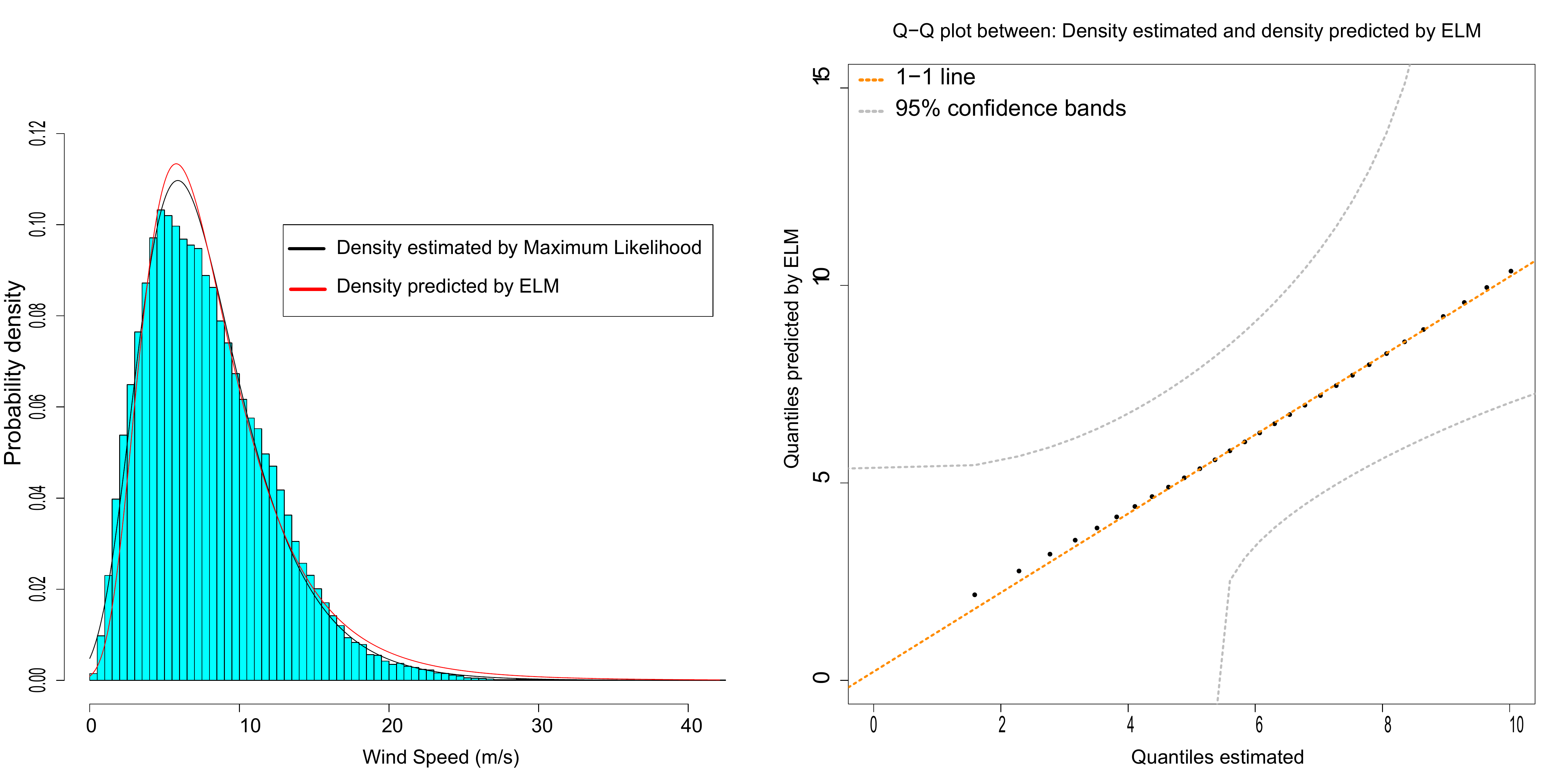}
\caption{Comparison between predicted and estimated distribution for $Jungfraujoch$ ($3580$ meters) taken from the testing set.}
\label{fig:9}  
\end{figure}

\section{Results and discussions}
\label{sec:5}
The presented work allows us to visualize better each parameter of the GEV distribution. Moreover, the produced maps are coherent with the topographical information in Switzerland. The very important parameter of the GEV is the shape $\xi$, it defines which subfamily fits the used data. A visualization of the shape on the map of Switzerland can help to answer to the question (fig. 10): which subfamily is used in a random place?\\ 
Such information is important for making useful analysis regarding risk assessment, and renewable energy produced by wind speed.

\begin{figure}
\centering
\includegraphics[scale=0.23, angle=-90]{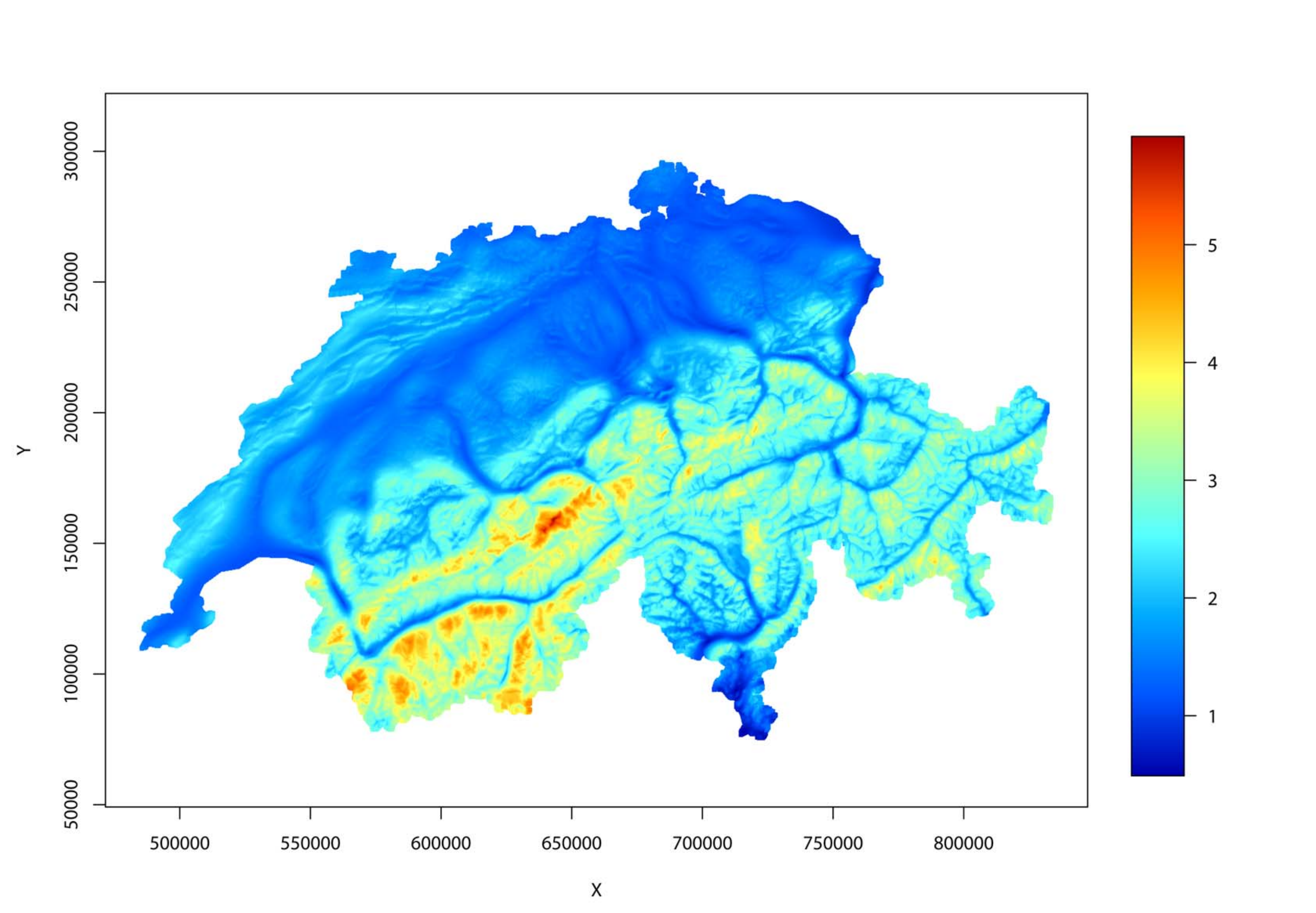}
\includegraphics[scale=0.23, angle=-90]{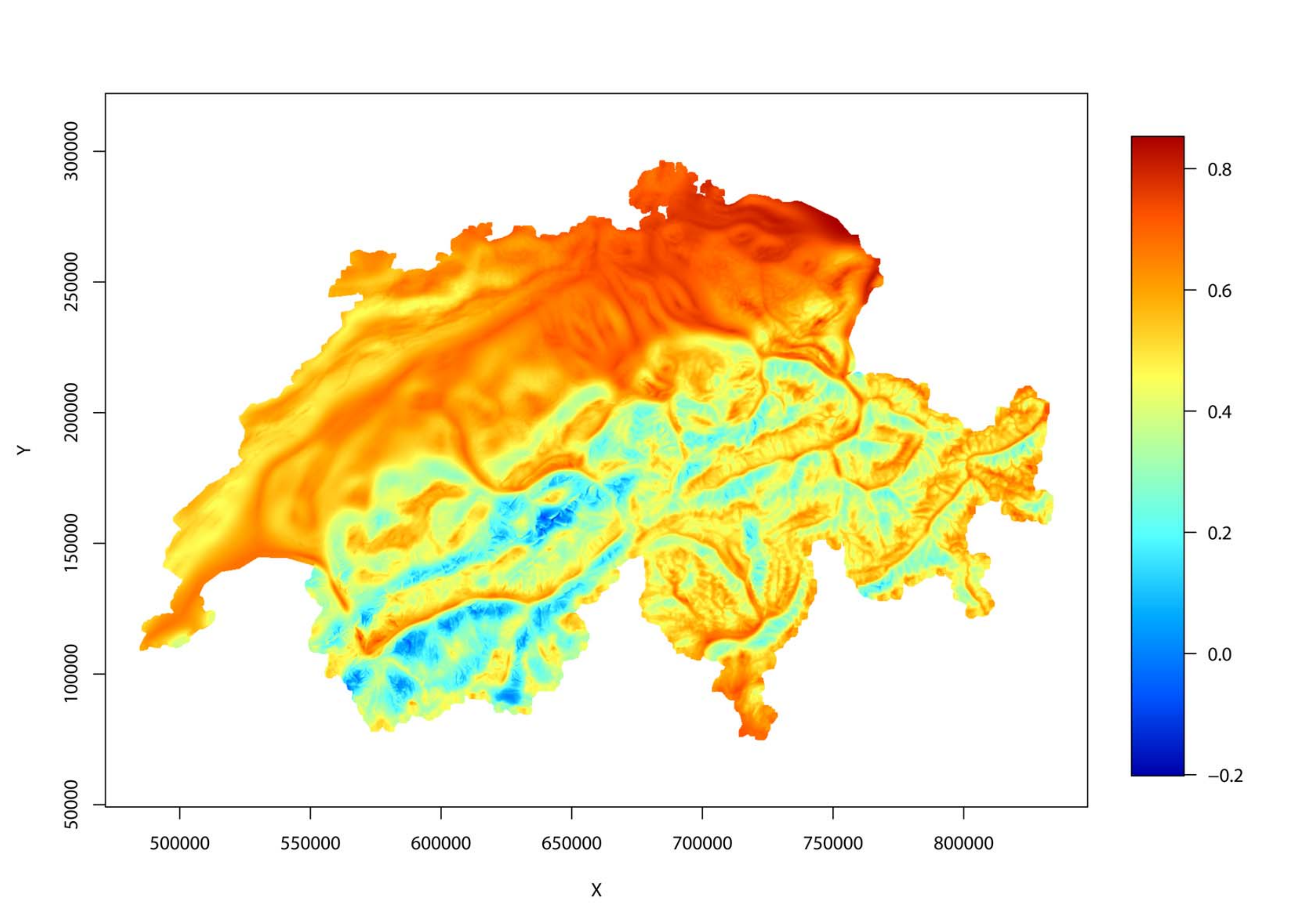}
\caption{On the left: Visualisation of the location parameter of GEV distribution. On the right:  The shape parameter, the dominated subfamilies in this case of study are more or less the Gumbel and the Fréchet subfamilies.}
\label{fig:10}  
\end{figure}

Furthermore, the advantage of knowing the extreme distribution allows us to predict the probability for a given wind speed and vice versa, fig. 11 presents probability map that a wind speed exceeds $15$ m/s.

\begin{figure}
\centering
\includegraphics[scale=0.45, angle=-90]{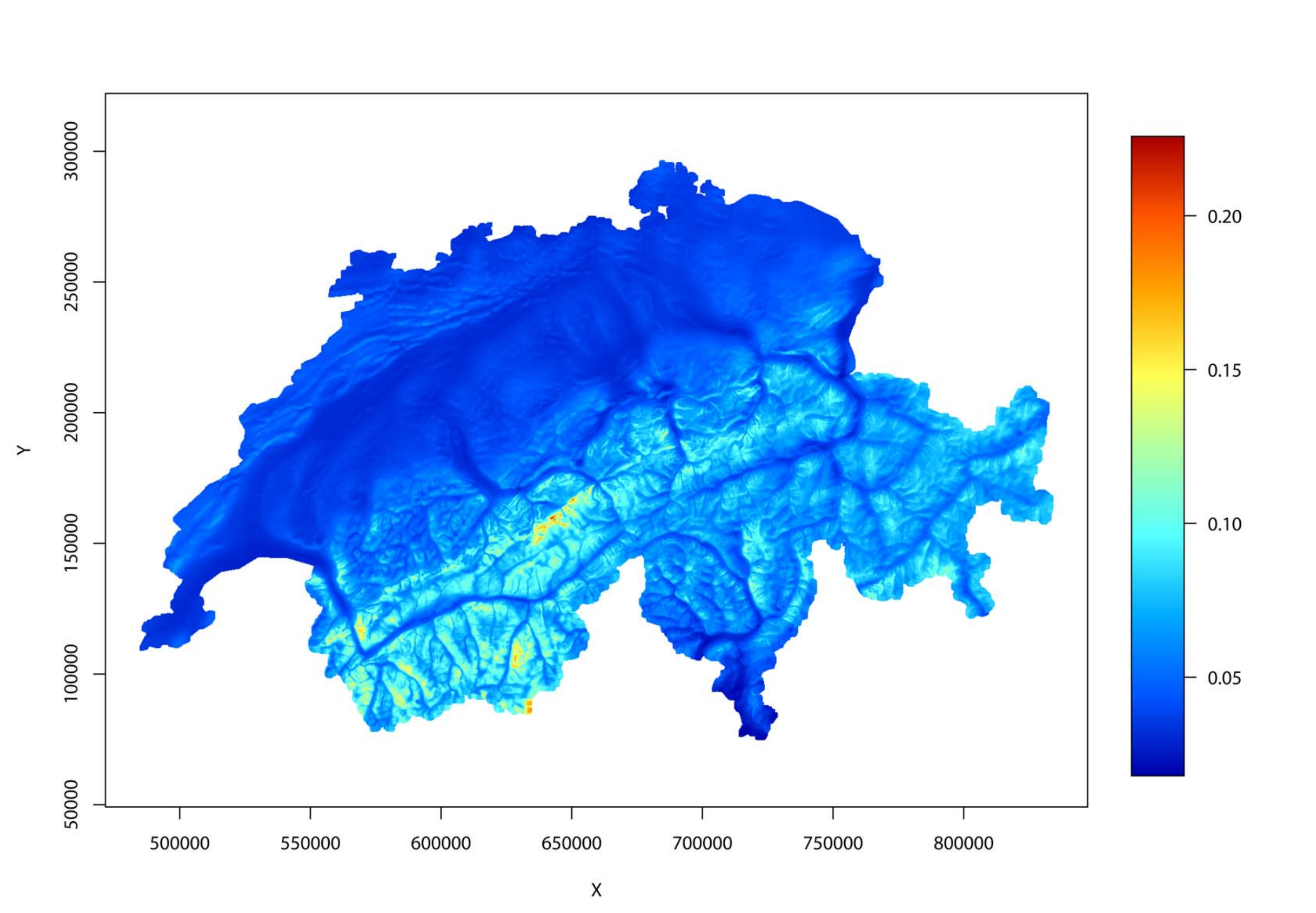}
\caption{Probability map for wind speed greater than or equal to 15 m/s.}
\label{fig:11}  
\end{figure}

Finally, this work used ELM to model each parameter of the GEV. This modelling approach deals with the repetition of prediction $20$ times to consider the randomness of ELM when it generates the weights. In order to compare the values given by ELM, fig. 12 shows different densities generated by the minimum values, the maximum, and the mean of the predicted parameters after the $20$ repetition. This difference is checked by using QQ-plot between the predicted values and the testing data. As expected, ELM shows its efficiency to model, by the insignificant different between results. However, repeat ELM several times improves the quality of results, and helps to obtain optimal model.

\begin{figure}
\centering
\includegraphics[scale=0.45]{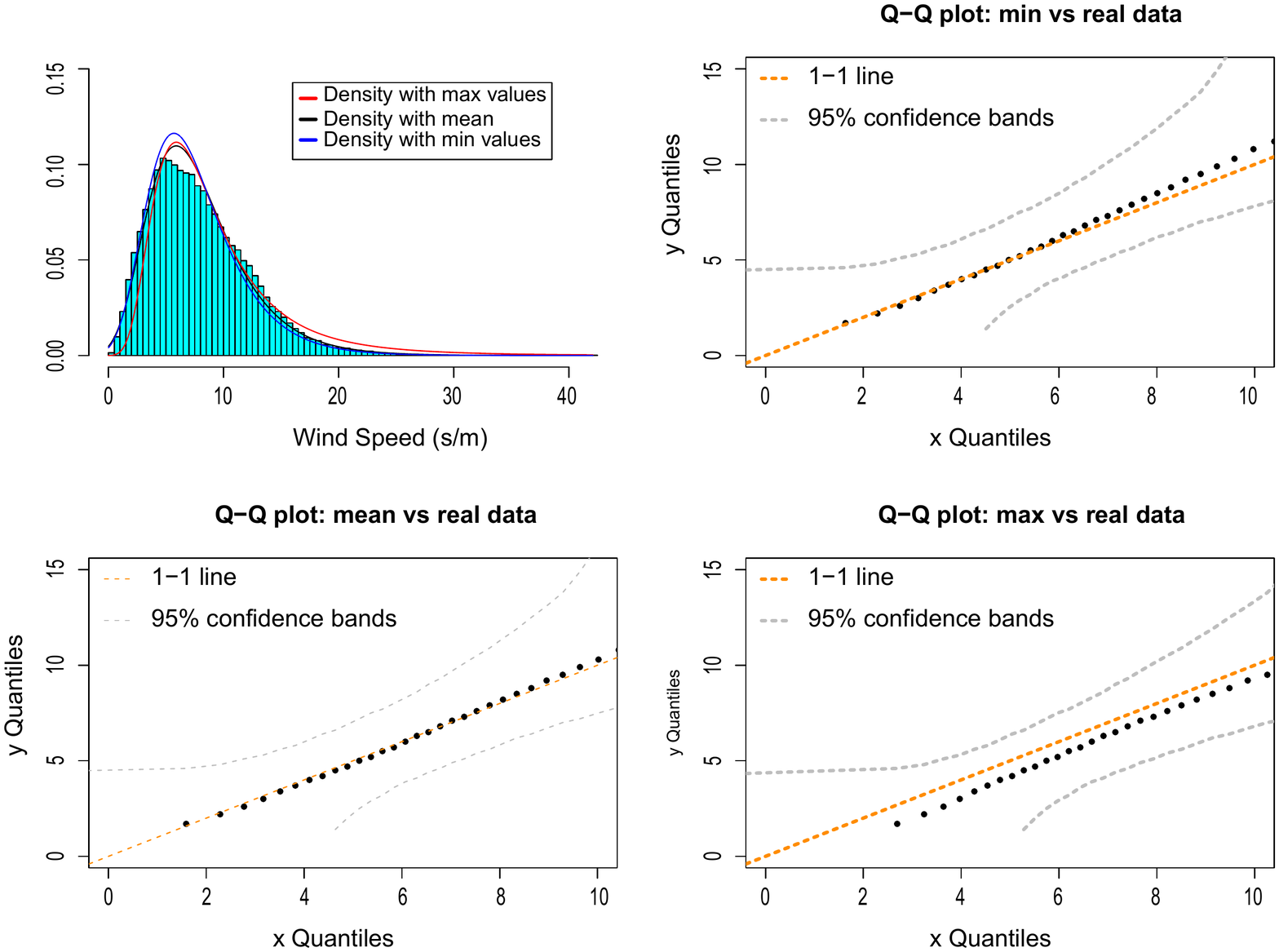}
\caption{Comparison between different results given by ELM.}
\label{fig:12}  
\end{figure}

One of the most important result in this work is the possibility to predict extreme wind speed in new places without measuring stations. Furthermore, the comparison between the three proposed distributions offers more precision to the study.
\section{Conclusion}
\label{sec:6}
Wind energy remains the most attractive resource for providing sustainable power. The research presented here gives an initial overview of wind speed data, and a global idea about extremes of this phenomenon in Switzerland. Such modelling can be useful in developing intelligent decisions for wind-powered electrical generators. Spatial modelling of distributions can be used to optimize existing network and to propose new places for the aeolian energy production.
The developed methodology provides more efficiency. The combination of parametric estimation and extreme learning machine offers more rapidity to obtain good results. Furthermore, the results are focused on extremes which is important for natural hazards and risk assessments. 
This methodology could be applied to other extreme environmental phenomena, e.g. precipitation.\\
Further developments could be in application of this methodology for spatio-temporal environmental data and quantification of the uncertainties. 

\section*{Acknowledgements}
The authors are grateful to Jean Golay and Michael Leuenberger for many fruitful discussions about machine learning and extreme values.\\
The authors thank MeteoSuisse for giving access to the data via IDAWEB server.\\
This research was partly supported by the Swiss Government Excellence Scholarships for Foreign Scholars.
\nocite{*}

\bibliography{xampl}
\bibliographystyle{elsarticle-num}


\end{document}